\documentclass{PoS}

\usepackage{amssymb}
\usepackage{times}
\usepackage{mathptmx}
\usepackage{fontenc}
\usepackage{psfig}
\usepackage{graphicx}
\usepackage{cite}

\title{Chemical Evolution of the Carina Dwarf Spheroidal}

\ShortTitle{Chemical Evolution of the Carina Dwarf}

\author{\speaker{Kate Pilkington}\\%
       University of Central Lancashire\\
       E-mail: \email{kpilkington@uclan.ac.uk}}

\author{Brad K. Gibson\\
        University of Central Lancashire\\
        E-mail: \email{bkgibson@uclan.ac.uk}}

\abstract{We explore a range of chemical evolution models for the Local 
Group dwarf spheroidal (dSph) galaxy, Carina.  A novel aspect of our 
work is the removal of the star formation history (SFH) as a `free 
parameter' in the modeling, making use, instead, of its colour-magnitude 
diagram (CMD)-constrained SFH.  By varying the relative roles of 
galactic winds, re-accretion, and ram-pressure stripping within the 
modeling, we converge on a favoured scenario which emphasises the 
respective roles of winds and re-accretion. While our model is 
successful in recovering most elemental abundance patterns, comparable 
success is not found for all the neutron capture elements.  Neglecting 
the effects of stripping results in predicted gas fractions 
approximately two orders of magnitude too high, relative to that 
observed.}

\FullConference{XII International Symposium on Nuclei in the Cosmos\\
   August 5-12, 2012\\
    Cairns, Australia}

\begin{document}

\section{Introduction}

\begin{figure}[b]  
\begin{center}
\hspace{0.25cm}
\psfig{figure=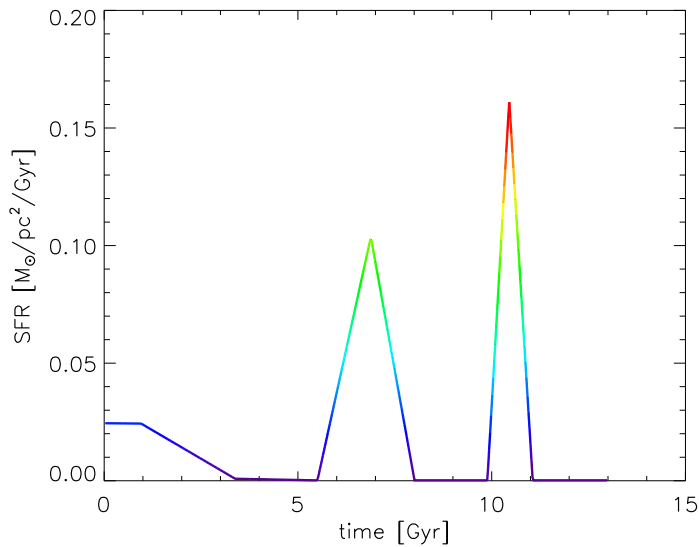,width=10.cm}
\caption{Empirically-derived star formation history of the Carina dwarf 
spheroidal\cite{Dol05}, employed as a `fixed' input to the chemical 
evolution model, colour-coded according to star formation rate (see 
Fig.~3).}
\label{sfh}
\end{center}
\end{figure}

Local Group dwarf galaxies are critical training sets for all aspects of 
Galactic Archaeology and high-redshift galactic chemical evolution. They 
are the only systems in the Universe for which the otherwise highly 
uncertain (essentially unconstrained) star formation history (SFH) is 
not an unknown. If one cannot demonstrate an ability to model the 
chemical evolution of such constrained systems, it limits one's 
confidence in the ability to do so for (the many more distant and 
numerous) unresolved stellar populations. We present models of the Local 
Group dwarf spheroidal (dSph) Carina, as a case study in our probe of 
the efficacy of galactic chemical evolution. We focus on the use of 
CMD-derived star formation histories \cite{Dol05} coupled to our 
chemical evolution package \tt GEtool\rm\cite{FenGib}. We sample a range 
of inflow and outflow parameterisations, concentrating on Carina due to 
the unprecedented quality of its spectroscopic data \cite{Venn}. Carina 
shows three main sequence turn-offs in its CMD, corresponding to the 
stellar populations associated with distinct star formation episodes 
(Fig.~\ref{sfh}), the majority of which belong to the intermediate-age 
burst.

\tt GEtool \rm tracks the time evolution of 107 isotopes (spread over 45 
elements); as in our preliminary work on the Sculptor dwarf 
\cite{Fen06}, we employ SNeII yields \cite{WooWea}, SNeIa 
yields\cite{Iwam99}, and post-processed neutron-capture yields from AGB 
stars \cite{Lug03}.  The underlying yields from low- and 
intermediate-mass stars has been updated, to reflect recent developments 
in the field\cite{Kar}. \tt GEtool \rm allows consideration of feedback 
from supernova, outflowing galactic winds, parameterised ram-pressure 
stripping, re-accretion of stripped/wind ejecta, and infall of fresh 
fuel for future generations of star formation. We adopt a Kroupa, Tout 
\& Gilmore\cite{Kroupa93} initial mass function (IMF) for the work 
described here.

\section{Inflows and Outflows}

\begin{figure}[h]  
\begin{center}
\hspace{0.25cm}
\psfig{figure=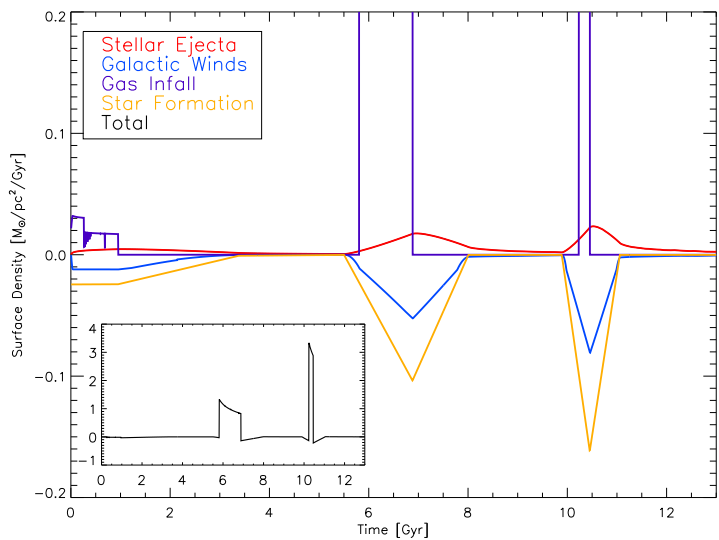,width=10.cm}
\caption{The evolution of the mass surface density of gas. In red is 
shown the material returned to the system from dying stars, as a 
function of time; blue represents the amount of gas removed by the 
galactic winds; yellow represents the amount of gas used up in star 
formation; purple represents the gas infalling to the system. The inset to 
the panel shows the sum of the four sources of gas - clearly, infalling 
material dominates over the other sources ($\sim$10:1 relative to, for 
example, outflowing wind material).}
\label{inout}
\end{center}
\end{figure}

We first show the inferred temporal evolution of the total gas surface 
density of Carina (inset panel within Fig.~\ref{inout}). We can 
sub-divide this `total' into four primary sub-components; gas resulting 
from stellar ejecta (red), that taking part in a galactic wind/outflow 
(blue), that associated with fresh infall of star formation `fuel' 
(purple), and that being `lost' to star formation at any given time 
(yellow).  The dominant role of gas infall (purple) is readily apparent; 
in some sense, the primary novel aspect of our modeling is that said 
infall is parameterised (or `controlled') to ensure the model adheres 
strictly to a Kennicutt star formation law of the form 
$\psi(t)=0.05~\sigma(t)^{1.4}$~M$_\odot$~pc$^{-2}$~Gyr$^{-1}$ (where 
$\psi$ is the star formation rate and $\sigma$ is the gas surface 
density).

In the absence of exceedingly efficient Type~Ia supernovae 
(SNeIa)-driven outflows and/or ram pressure stripping, the predicted 
final gas fraction of the model would be $\sim$90\% - i.e., roughly two 
orders of magnitude higher than observed. Parameterised ram pressure 
stripping \cite {Pas12} should allow us to better recover the low gas 
fractions seen in dSphs in the vicinity of massive hosts, like the Milky 
Way. In the interim, simply stripping the most recently infallen fuel 
(least tightly bound) provides a suitable final fraction.

\section{Abundance Patterns}

\subsection{Alpha Elements}

In Fig~\ref{abs}, we show the abundance patterns of three 
$\alpha$-elements: magnesium, oxygen, and calcium. We find reasonable 
agreement with the observational data, particularly for oxygen and 
calcium. Our predictions for magnesium are less ideal, in large part we 
feel, due to the well-known issue concerning the underproduction of Mg 
from the SNeII models of Woosley \& Weaver \cite{WooWea}. Here, we have 
allowed for re-accreation of outflowing wind material; not doing so, 
within the context of our framework, leads to an underproduction of the 
global stellar metallicity. Our earlier models\cite{Fen06} suffered from 
a significant overproduction of sodium, but as shown clearly in Fig~3, 
this problem has been rectified naturally via the use of the newer AGB 
models of Karakas (2010).

\subsection{Neutron Capture Elements: r- and s-process}

\begin{figure}  
\begin{center}
\hspace{0.25cm}
\psfig{figure=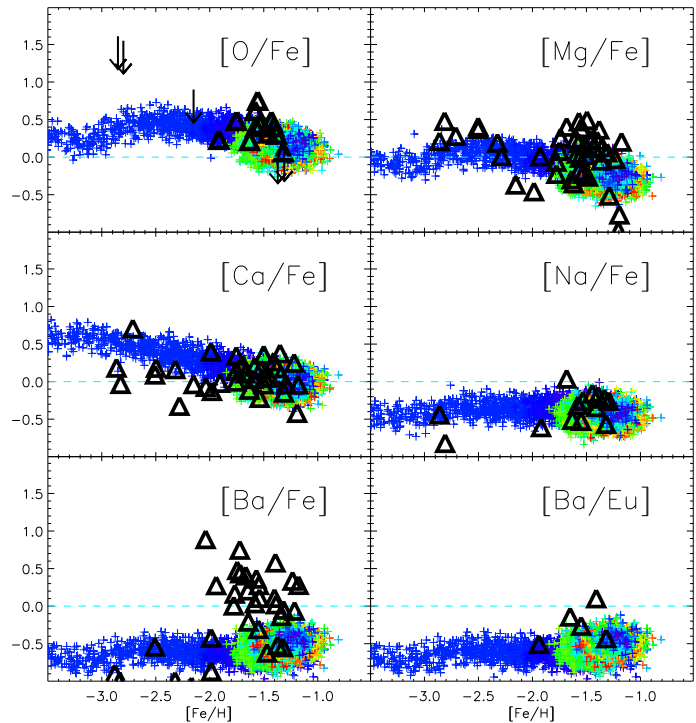,width=10.cm}
\caption{Predicted abundance patterns for the Carina dSph (small dots) 
coloured according to the star formation rate (recall, Fig~1), where red 
and green symbols correspond to times of higher star formation rates, 
while blue and purple correspond to periods of lower star formation. The 
black triangles correspond to the observational data of Venn et~al. 
\cite{Venn}; downward facing arrows are also from Venn et~al, but 
represent data for which only upper limits exist.}
\label{abs}
\end{center}
\end{figure}

Also in Fig~\ref{abs}, we show the predicted distributions of the heavy 
s-process element barium and the r-process element europium. At low 
metallicities, we capture the behaviour of the neutron capture elements 
well, but it is equally clear that we grossly underestimate the dynamic 
range in both the s- and r-process elemental patterns at higher 
metallicities.

\subsection{Metallicity Distribution Function (MDF)}

\begin{figure}  
\begin{center}
\hspace{0.25cm}
\psfig{figure=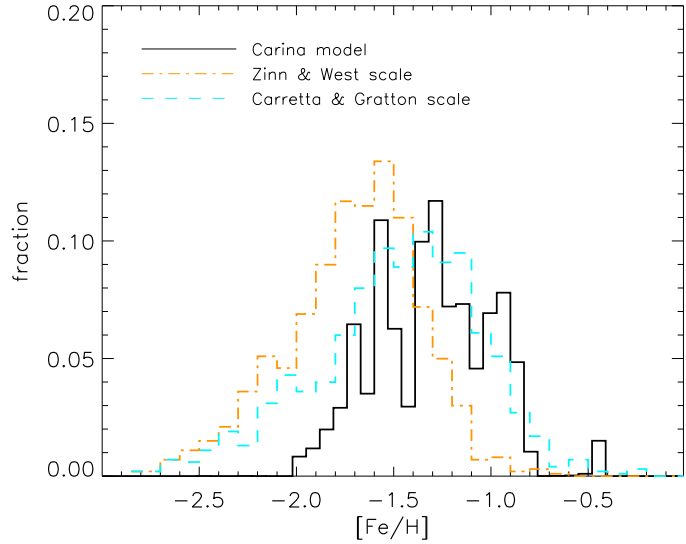,width=10.cm}
\caption{The predicted metallicity distribution function of the Carina 
dSph (black) compared with the observed MDF of Koch et~al. \cite{Koch06} 
(calibrated with two different metallicity calibrations: \cite{CaGr} in 
cyan, and \cite{ZiWe} in orange). Our model has been convolved with a 
$\sigma$=0.28~dex Gaussian to mimic the uncertainties associated with 
observational data.}
\label{mdf}
\end{center}
\end{figure}

In Fig~\ref{mdf}, we show the MDF of our fiducial Carina model (black 
histogram), alongside the observed MDF from \cite{Koch06} (derived using 
two different metallicity calibrations: \cite{CaGr}, in cyan, and 
\cite{ZiWe} in orange). The model MDF has been convolved with a 
$\sigma$=0.28~dex Gaussian, to reflect the quoted observational 
uncertainties\cite{Koch06}. One can see that the peak of our MDF matches 
well with the observed peak (cyan), although the current fiducial model 
admittedly suffers from a dearth of lower metallicity stars in the tail 
of the MDF.

\section{Conclusions}

\begin{enumerate}
\item{We show that an infall rate which is recovered by inverting the 
Kennicutt star formation law ($\psi\propto\sigma^{1.4}$) with a 
CMD-inferred star formation history can successfully match many of the 
chemical properties of the Carina dSph, without the need for additional 
`fine-tuning' of the basic chemical evolution properties.}
\item{Most elemental abundance patterns are consistent with those 
observed, save (primarily) for the neutron capture elements.}
\item{Without invoking some form of parameterised ram pressure 
stripping, we inevitably over-predict the final gas fraction by two 
orders of magnitude.}
\item{Extending our work to the cover the entire sample of Local Group 
dwarfs with CMD-inferred star formation histories, including samples 
such as those of LCID \break
(\tt http://http://www.iac.es/proyecto/LCID/\rm), 
is one of the next steps in this work.}
\end{enumerate}

\end{document}